\newcommand{\al}{\mbox{$\alpha $}}

\newcommand{\e}{\mbox{$e^{ik.X(z)}$}}
\newcommand{\qe}{\mbox{$e^{iq.X(0)}$}}
\newcommand{\pe}{\mbox{$e^{ip.X}$}}

\documentstyle[12pt]{article}
\newcommand{\kim}{\mbox {$ k_{1}^{\mu}$}}
\newcommand{\kom}{\mbox {$ k_{0}^{\mu}$}}

\newcommand{\kin}{\mbox {$ k_{1}^{\nu}$}}

\newcommand{\ktm}{\mbox {$ k_{2}^{\nu}$}}

\newcommand{\lpp}{\mbox {$e^{i\int _{c} \alpha (t)
k(t) \partial _{z} X(z+t) dt +ik_{0}X}$}}

\newcommand{\la}{\mbox{$ \lambda $}}
\newcommand{\be}{\begin{equation}}
\newcommand{\br}{\begin{eqnarray}}
\newcommand{\ee}{\end{equation}}
\newcommand{\er}{\end{eqnarray}}

\newcommand{\gvk}{\mbox {$ e^{i\sum _{n}k_{n}Y_{n}}$}}

\newcommand{\eln}{\mbox {$ e^{\sum _{n}\lambda _{-n}L_{n}}$}}

\newcommand{\dsi}{\mbox {$\frac{\partial \sigma}{\partial x_{1}}$}}

\newcommand{\dsq}{\mbox {$\frac{\partial \sigma}{\partial x_{n+m}}$}}
\newcommand{\dst}{\mbox {$\frac{\partial \sigma}{\partial x_{2}}$}}

\newcommand{\dds}{\mbox {$\frac{\delta}{\delta \sigma}$}}

\newcommand{\dsnm}
{\mbox {$\frac{\partial ^{2}\sigma}{
\partial x_{n}\partial x_{m}}$}}
\newcommand{\dsii}{\mbox {$\frac{\partial ^{2}\sigma}
{\partial x_{1}^{2}}$}}

\newcommand{\p}{\mbox {$ \partial$}}

\newcommand{\pp}{\mbox {$ \partial ^{2}$}}
\newcommand{\mup}{\mbox {$ \partial _{\mu}$}}

\newcommand{\ppp}{\mbox {$ \partial ^{3}$}}

\begin{document}
\title{On Covariant Derivatives and
Gauge Invariance in the Proper Time Formalism for String Theory}
\author{B. Sathiapalan\\ {\em
Physics Department}\\{\em Penn State University}\\{\em 120
Ridge View Drive}\\{\em Dunmore, PA 18512}}
\maketitle
\begin{abstract}
It is shown that the idea of ``minimal'' coupling to gauge fields
can be conveniently implemented in the proper time
formalism by identifying
the equivalent of a ``covariant derivative''.  This captures some of
the geometric notion of the gauge field as a connection.
The proper time equation
is also generalized so that the gauge invariances associated with
higher spin massive modes can be made manifest, at the free level,
using loop variables.  Some explicit examples are worked out illustrating
these ideas.
\end{abstract}
\newpage
\section{Introduction}

A proper understanding of the massive modes in string theory is essential,
not only from a conceptual standpoint, but also from a more practical,
computational standpoint.  In the first quantized (Polyakov) formalism,
where the massless modes are represented by marginal operators, these
massive modes are represented by irrelevant operators.  The renormalization
group equations of these theories and their connection with the
equations of motion of the string modes has been the subject of a lot of papers
[1-25].  Exact solutions of these equations have also been investigated
\cite{t3}.  The bulk of this work deals with massless modes.  The massive modes
are more difficult to deal with for two reasons:  First, being massive, one
has to deal with space time dependent fields in order to satisfy the
requirement that the corresponding vertex operators are marginal.  This
requirement arises because the usual $\beta $-function calculations are
conveniently done only for marginal operators.  Even if one is willing to deal
with space-time dependent fields, the calculations are complicated because
one typically has to sum an infinite number of diagrams to obtain non-trivial
interactions \cite{ds,amit}.  Second, the massive modes have higher spin and
there are many extra symmetries associated with them.  While the free
equations have been obtained \cite{bslv,bslv1}, the subject of interactions
has not been dealt with.  At the present time, one has to resort to string
field theory  \cite{sz,bp,w2,lpp,gj,z}for this. Indeed, the interaction of
electromagnetic and massive modes, in the limit of uniform electromagnetic
field strength,
has been studied using string field theory \cite{an}.

Our aim is to apply the proper time formalism \cite{bspt} to address
the problem of massive modes.
A prerequisite for understanding the masssive case is the massless case.
Some aspects of the massless case have been discussed in earlier papers
\cite{bsg,bsrg,bsz}.
In this paper we discuss the ``minimal'' interaction of electromagnetism with
other modes.
We are
interested in the general case of non-uniform fields, but with the
restriction that they be
close to their
mass shell.
In the case of massive higher spin modes the minimal interaction
is not gauge invariant by itself
 and one has to include the direct coupling to the
field strength for consistency.
However we will not discuss the direct coupling to the field strength
in this paper.  Instead we will
concentrate on getting a better understanding of how to implement
the minimal interaction.  This was worked out in some approximation
in \cite{bsg}.  In this paper we show that something like
the ``covariant derivative'',
an indispensable tool in gauge theories, can also be introduced here.
In \cite{bsg} it was observed that in deriving the covariant Klein Gordon
equation each term arose as a surface term.  This will become manifest
in the more general derivation given here.

There is another issue that needs to be resolved before one can begin to
discuss the interaction of massive modes in the proper time formalism.
We need to generalize the formalism to keep track of the extra
gauge symmetries associated with the massive modes.  In \cite{bslv,bslv1}
the gauge invariant free equations of massive modes were worked out
by requiring that a representation of the string field, called a
``loop variable'', not have any anomalous dependence on the Liouville mode.
This technique has some advantages, the principle one probably being
simplicity. We transcribe this into the framework
of the proper time equation, which is convenient for working
out the interacting equations of motion.
This is the second main result of this paper.
We also give an explicit calculation by
way of illustration.  It is also interesting to note that, in this form,
the proper time equation looks very similar to the equation written
down by Witten in his background independent formalism \cite{w,wl,s}.

In section II we describe the minimal coupling of electromagnetism
and the covariant derivative, and illustrate it with the coupling to the
tachyon.  In section III we generalize the proper time formalism to
higher spin massive modes and show how the loop variable formalism
can be incorporated.  Section IV contains some conclusions.

\newpage
\section{Minimal Coupling and the ``Covariant Derivative''}

The proper time equation, in its simplest form for the tachyon is:
\be                  \label{pt}
\left. \int d^{D}k \{ \frac{d}{d\ln z} z^{2} <V_{p}(z) V_{k}(0)>\}
\right | _{\ln  z=0}
\phi (k) =0 .
\ee
Here $V_{p}=\pe$.
We will use this equation to illustrate the minimal coupling.  The
expectation value in (\ref{pt}) is calculated using the sigma model action
appropriate to the problem.  In the present case we will use the
usual Polyakov action supplemented by an electromagnetic field
interaction given by:
\be       \label{2.3}
\Delta S \, = \, \int dz \int d^{D}p A_{\mu}(p) \p _{z} X^{\mu} e^{ip.X(z)}
\ee
The ``minimal'' part of the interaction can be isolated by using the
following identity \cite{bsfc,bsz}:
\be     \label{id}
A_{\mu}(k)\p _{z} X^{\mu} e^{ikX(z)} =
\int _{0}^{1}d \alpha \p _{z} (A_{\mu} X^{\mu} e^{i\alpha kX})
+i\int _{0}^{1} d \alpha \alpha
k_{[\mu}A_{\nu]}(X^{\mu}\p _{z} X^{\nu} e^{i\alpha kX})
\ee
When the RHS of (\ref{id}) is substituted in (\ref{2.3}), the first term
represents the minimal interaction and the second term represents the
direct coupling to the field strength $F_{\mu \nu}$.  In this paper
we will discuss only the minimal coupling.  We have to evaluate
\be    \label{2.4}
\left. \int d^{D}q \{ \frac{d}{d\ln z} z^{2} <e^{ikX(z)} e^{i\Delta S}
e^{iqX(0)}>\}
\right | _{\ln  z=0}
\phi (q) =0 .
\ee
If we insert the identity (\ref{id}) in $\Delta S$ we get (keeping only
the first term of (\ref{id}))
\be  \label{2.5}
\left. \int d^{D}q \{ \frac{d}{d\ln z} z^{2} <e^{ikX(z)}
e^{i\int _{0}^{1}d \alpha \int _{0}^{z}dw
\partial _{w} (A_{\mu}(p) X^{\mu} e^{i\alpha pX})  }
e^{iqX(0)}>\}
\right | _{\ln  z=0}
\phi (q) =0 .
\ee

We have suppressed the integral over $p$.
The integral over $w$ gives a boundary term and we get
\be     \label{2.6}
\left. \int d^{D}q \{ \frac{d}{d\ln z} z^{2}
<e^{i(k_{\mu}+ \int _{0}^{1}d \alpha A_{\mu}(p)e^{i\alpha pX}) X^{\mu} (z) }
e^{i(q_{\mu}-\int _{0}^{1}d \alpha A_{\mu}(p)e^{i\alpha pX})X^{\mu}(0)}>\}
\right | _{\ln  z=0}
\phi (q) =0 .
\ee

One can easily see the effect of a gauge transformation $A_{\mu}
\rightarrow A_{\mu}(k) + ik_{\mu} \Lambda$:
\[
\int d^{D}k \int_{0}^{1} d \alpha A_{\mu} X^{\mu} e^{i\alpha kX(0)} \rightarrow
\]
\be     \label{2.7}
\int d^{D}k \int_{0}^{1} d \alpha A_{\mu} X^{\mu} e^{i\alpha kX(0)}
-i \int dk \Lambda (k) e^{ikX(0)} +i\int dk \Lambda (k)
\ee
On inserting (\ref{2.7}) into the second exponential factor in (\ref{2.6})
we find that the second term (of (\ref{2.7}))
is cancelled by the variation $\phi \rightarrow \phi
e^{i\Lambda (X(0))}$.  The third piece is a space-time independent
constant, i.e. a global phase, which is cancelled by a corresponding term
from the first exponential factor at the point $z$.  Thus (\ref{2.6})
transforms by a phase factor $e^{i\Lambda (X(z))}$, which shows that
the equation of motion that one obtains from (\ref{2.6}) is guaranteed
to be covariant and can therefore be
expressed in terms of covariant derivatives. We will see this
in an explicit calculation.
In appearance, the exponent in (\ref{2.6}) viz.
\be     \label{2.8}
(k_{\mu}+ \int _{0}^{1}d \alpha A_{\mu}e^{i\alpha pX})X^{\mu}
\ee
looks like $ X^{\mu}D_{\mu}$.
Of course at leading order
this is guaranteed by the gauge invariance arguments given above and the
fact that the $A$-independent part is $ X^{\mu} \mup$.  But {\em a priori}
it need not
be true at the next order where there are several Lorentz invariant
combinations possible.
What is interesting is that, as we shall see, this is exact
even at the next-to-leading order.
  From (\ref{2.5}) and (\ref{2.6})
we can also immediately see an explanation for the
fact observed in \cite{bsg} that in deriving the Klein Gordon equation
only surface terms contribute.

Let us now turn to the actual evaluation of (\ref{2.6}).
We bring down one power of $A_{\mu}$ from each of the exponents in
(\ref{2.6}) to get:
\be     \label{2.9}
<\e [1+ i\int _{0}^{1}d \alpha A(p)X(z)e^{i\alpha pX(z)}]
\qe [1-i\int _{0}^{1}d \beta A(p')X(0)e^{i\alpha p'X(0)}]>
\ee
There are two points to note here.  We have refrained from expanding
$e^{ikX}$ as a power series in $k.X$.  This is only to make explicit
the appearance of momentum conservation delta functions.  As far as
the actual algebra is concerned, one can just as well expand the
exponential in a power series, keeping as many terms as necessary,
and the answer would be the same.  The second point concerns the
integral over $\alpha$ and the presence of the exponential
$e^{i\alpha p.X}$.  This has a momentum $\alpha p$. But we would like
$A_{\mu}(p)$ to be associated with a vertex operator of momentum $p$.
Therefore we will write
\be     \label{2.10}
e^{i\alpha pX}\, = \, e^{ipX +i(\alpha -1)pX} \, = \, e^{ipX}[1+
i(\alpha -1)pX + ...]
\ee
Thus at each order in $(\alpha -1)p.X$ we have a vertex operator
of momentum $p$.  Of course for the purpose of doing the algebra
$\alpha $ can be kept in the exponent, as long as we remember that it
actually represents a sum of vertex operators, each of momentum $p$.

Let us evaluate (\ref{2.9}).  Using \footnote{All the calculations
in this section are valid for point particles also.
All we have to do is
replace (\ref{2.11}) by $<X(T)X(0)> \, \, = \, iT/2 $}
\be     \label{2.11}
<X^{\mu} (z) X^{\nu} (0)> \, = \, -\ln z \delta ^{\mu \nu}
\ee
we get
\[
e^{k.q \ln z} \delta (k+q)
+ \int _{0}^{1} d \alpha A(p).q \ln z e^{(k+\alpha p).q\ln z} \delta (k+p+q)
\]
\[
-\int _{0}^{1}d \beta A(p).k \ln z e^{k.(q+\beta p)} \ln z \delta (k+p+q)
\]
\be     \label{2.12}
+\int _{0}^{1} d \alpha \int _{0}^{1} d \beta A(p).A(p')
e^{k+\alpha p ).(q+\beta p) \ln z} \delta (k+p+p'+q)
\ee
Simplifying, we get on substituting into the proper time equation
(\ref{2.4})
\be     \label{2.13}
-(q^{2}-2) \delta (q+k) + A(p).(2q+p) \phi (q) \delta (q+p+k) +
A(p).A(p')\phi (q) \delta (k+p+p'+q) \, = \, 0
\ee
In coordinate space this is
\be     \label{2.14}
(D_{\mu}D^{\mu}+2) \phi \, = \, 0
\ee

Heuristically, if we think of expression (\ref{2.8}) as standing for
$X^{\mu} D_{\mu}$, then to lowest non-trivial order (\ref{2.6})
becomes
\be     \label{2.15}
<(1+X^{\mu}(z)D_{\mu})(1+X^{\nu}(0)D_{\nu})>
\ee
Using (\ref{2.11}) we get from the proper time equation, the result
(\ref{2.14}).

Let us turn to the next order to see if the heuristic identification
of (\ref{2.8}) with $X^{\mu} D_{\mu}$ has any value.  If correct,
one should obtain:
\be     \label{2.16}
<\frac{D^{\mu}D^{\nu} X^{\mu}X^{\nu}}{2!}
\frac{D^{\rho}D^{\sigma} X^{\rho} X^{\sigma}}{2!}>
\ee
\be     \label{2.17}
= \frac{D^{\mu}D^{\nu}D^{\rho}D^{\sigma}}{4}
(\delta ^{\mu \rho}\delta ^{\nu \sigma} +
\delta ^{\mu \sigma}\delta ^{\nu \rho}) (\ln z )^{2}
\ee
\be     \label{2.18}
= \frac{( D^{\mu}D^{\nu}D_{\nu}D_{\mu}+D^{\mu}D^{\nu}D_{\mu}D_{\nu})}{4}
(\ln z)^{2}
\ee

{\em A priori} there are three possible structures at this order:
$D^{2}D^{2}$, $D^{\mu}D^{2}D_{\mu}$ and $D^{\mu}D^{\nu}D_{\mu}D_{\nu}$.
If we let $D \equiv \p - iA$, so that $[D_{\mu},D_{\nu}] \, = \,
-iF_{\mu \nu}$, the following are easily proved:
\be     \label{2.18.1}
 D^{\mu}D^{\nu}D_{\nu}D_{\mu} =
 D^{2}D^{2} + i \mup F_{\mu \nu} D^{\nu} + F_{\mu \nu}F^{\mu \nu}
\ee
\be     \label{2.18.2}
 D^{\mu}D^{\nu}D_{\mu}D_{\nu} =
 D^{2}D^{2} + i \mup F_{\mu \nu} D^{\nu} + \frac{1}{2} F_{\mu \nu}F^{\mu \nu}
\ee

Let us now check if (\ref{2.18}) is correct.
At second order one has to evaluate:
\[
<\{ 1+i(k+\int _{0}^{1}d \alpha A(p)e^{i\alpha pX(z)})X(z)
+\frac{i^{2}}{2!}[(k+\int _{0}^{1}d \alpha A(p)e^{i\alpha pX(z)})X(z)]^{2}\}
\]
\be     \label{2.18.3}
\{ 1+i(q-\int _{0}^{1}d \beta A(p')e^{i\beta pX(0)})X(0)
+\frac{i^{2}}{2!}[(q-\int _{0}^{1}d \beta A(p')e^{i\beta pX(0)})X(0)]^{2}\} >
\ee

First of all, it is easy to see that the term with four derivatives and the
term with four $A$'s are both consistent with $1/2 D^{2}D^{2}$ and therefore
(using (\ref{2.18.1}) and (\ref{2.18.2})) with (\ref{2.18}).  Next,
let us calculate the piece that is linear in $A$.  One obtains
\[
-A(p).q[(p+q)^{2}+q^{2}]-1/2 A(p).p[(p+q)^{2}+q^{2}]
\]
\be     \label{2.18.4}
+1/2[A(p).qp^{2}-A(p).pp.q]
\ee
Of the three terms in the above equation, the first two correspond
to the linear (in $A$) part of $1/2 D^{2}D^{2}$.  The third term
is $i/2 \mup F_{\mu \nu} \p ^{\nu}$ which is the linear (in $A$)
part of $i/2\mup F_{\mu \nu} D^{\nu}$.  Thus the result,
(\ref{2.18.4}), for the linear piece,
is consistent with
(\ref{2.18}). However it is also consistent with any linear
combination of the two terms, (\ref{2.18.1}) and (\ref{2.18.2}),
such that the
sum of their coefficients add up to 1/2.

To determine the precise
combination, we consider terms quadratic in $A$.  Let us concentrate,
for definiteness on terms of the type $A(p).p'A(p').p$ and
$A(p).A(p')p.p'$.  One finds from (\ref{2.18.3}):
\[
\frac{3}{4}A(p).p'A(p').p + \frac{1}{4}A(p).A(p')p'.p
\]
\be
= \frac{3}{4}[A(p).p'A(p').p - A(p).A(p')p'.p] + A(p).A(p')p'.p
\ee
The term in square brackets corresponds to
$3/8F_{\mu \nu}F^{\mu \nu}$ and the
second term is the contribution from $1/2D^{2}D^{2}$.  This determines
(using (\ref{2.18.1})and(\ref{2.18.2})) that it is precisely the linear
combination given in (\ref{2.18}) that is obtained.

We thus conclude that the heuristic identification of (\ref{2.8})
with $X^{\mu}D_{\mu}$ is correct, at least to this order.  If this is
true to all orders, then we have an easy way of writing down the result
at higher orders in $\ln z$ without the need to actually do
a detailed calculation.  It would certainly be interesting to know if this
is true to all orders.

Thus we have shown in this section how one can isolate the minimal
electromagnetic coupling in a simple way in the proper time equation.
One of the advantages of this ($\sigma$ model) formalism is that
it retains the
(space-time) geometric notion of the gauge field being a connection.
It should be possible to do something like this for the massive
modes as well.  In the next section we turn to the (free) massive modes.
\newpage
\section{Proper Time Equation for Massive Modes}
\setcounter{equation}{0}

In this section we will generalize the proper time equation
to make it covariant under the
extra gauge symmetries.  Reguiring that the coefficient of $\ln z$ vanish
is equivalent to imposing a dimensionality on the vertex operator, which
is the condition $L_{0}=1$.  For spin 1 and higher, we need to impose
further conditions of the form $L_{n}=0 \, ;\forall n>0$,  on the vertex
operators.  One thus starts by imposing
a linear combination of these constrains.  If the resulting equation
has the required gauge symmetries, one can then impose all the constraints as
gauge choices.  To this end let us consider the following object:
\be     \label{3.1}
\left. <\oint dt \lambda (t) T_{zz} (z+t) V_{I}(z) V_{J}(0)> \right |
_{\ln z =0} \Phi ^{J} (0)
\ee
with
\be     \label{3.2}
\lambda (t) = \lambda _{0}t + \lambda _{-1}t^{2} + \lambda _{-2} t^{3} +...
\ee
We will require that the coefficient of an appropriately chosen
linear combination of $\lambda _{-p}$ is zero.  Note that we have
evaluated (\ref{3.1}) at  $\ln z=0$ (or $\ln (z/a) =0$ where $a$ is a short
distance cutoff).  Note also that
$\left. <V_{I}(z) V_{J}(0)> \right | _{z=a}$ is the
Zamolodchikov metric \cite{zam}.

Before we apply it to the spin-2 case, as a warm up
exercise let us apply it to derive the covariant Klein Gordon equation.
Thus we consider:
\[
<\oint dt \lambda (t)\frac{1}{2} \p _{z} X(z+t) \p _{z}X(z+t)
\e
\]
\be     \label{3.3}
\left.
e^{i\int _{0}^{1}d \alpha \int _{0}^{z}dw
\partial _{w} (A_{\mu} X^{\mu} e^{i\alpha pX})  }
e^{iqX(0)}>
\right | _{\ln  z=0}
\phi (q) .
\ee
We have used the identity (\ref{id}) to isolate the minimal
interaction of the photon.  The momentum integrals are suppressed
for convenience. Doing the trivial $w$ integral gives:
\[
<\oint dt \lambda (t)\frac{1}{2} \p _{z} X(z+t) \p _{z}X(z+t)
e^{i(k_{\mu}+  \int _{0}^{1}d \alpha A_{\mu}e^{i\alpha pX}) X^{\mu} (z) }
\]
\be  \label{3.4}
\left.
e^{i(q_{\nu}-\int _{0}^{1}d \alpha A_{\nu}e^{i\alpha pX})X^{\nu}(0)}>
\right | _{\ln z=0} \phi (q)
\ee
We can expand the exponent just as before, to get
\[
<\oint dt \lambda (t)\frac{1}{2} \p _{z} X(z+t) \p _{z}X(z+t)
(\e + i \int _{0}^{1}d \alpha A_{\mu}(p)X^{\mu}(z)e^{i(k+\alpha p)X(z)}
\]
\be     \label{3.5}
\left.
 +\int _{0}^{1}d \alpha  \int _{0}^{1}d \beta A_{\mu}(p) A_{\nu}(p')
X^{\mu} (z) X^{\nu} (z)e^{i(k+\alpha p+ \beta p')X(z)}) \qe >\right |
_{\ln z =0} \phi (q)
\ee
Note that unlike what we did in Section II , we have kept two powers
of A from the first exponential and none from the second.  Any other
contribution is higher order in $\ln z$.  We thus get
\[
-\oint dt \frac{\lambda (t)}{t^{2}} <(\frac{k^{2}}{2}\e
+ \int _{0}^{1}d \alpha A(p).(k+\alpha p)(z)e^{i(k+\alpha p)X(z)}
\]
\be     \label{3.6}
\left.
-\frac{1}{2}\int _{0}^{1}d \alpha  \int _{0}^{1}d \beta
A(p).A(p')e^{i(k+\alpha p +\beta p')X(z)}) \qe > \right | _{\ln z=0}
\ee
\[
= \, \lambda _{0} [\frac{k^{2}}{2} \delta (k+q) +(k+\frac{p}{2}).A(p)
\delta (k+p+q)
\]
\be     \label{3.7}
- A(p).A(p')\delta(k+p+p'+q)] \phi (q)
\ee
As explained in the last section, in the above calculation,
$e^{i(k+\alpha p)X(z)}$ is being
interpreted as
\be     \label{3.8}
e^{i(k+p)X}(1+i(\alpha -1)pX+...)
\ee
Setting the coefficient of $\lambda _{0} $ to zero in (\ref{3.7})
gives the
(covariant) Klein Gordon equation.
\footnote{Since we have to set $L_{0}=1$,
(and not $L_{0}=0$) we should subtract
$\lambda _{0} e^{ikX}$ from (\ref{3.6}).  This will give the tachyon
its mass.}

We turn now to the modes with gauge invariances.  We will use
the loop variables that were introduced for this purpose in \cite{bslv}.
We give below a brief review:
The ``loop variable'' of \cite{bslv} describes all the modes of the string
and is the following:
\be     \label{3.9}
\lpp
\ee
which can be rewritten as
\be     \label{3.10}
e^{i(k_{0}Y+k_{1}Y_{1}+k_{2}Y_{2} +...+k_{n}Y_{n}+...)}
\ee
The $k_{n}$ are defined by
\be     \label{3.11}
k(t)=k_{0}+
\frac{k_{1}}{t}+
\frac{k_{2}}{t^{2}}+...
\ee
The $k_{i}$ define the fields as follows:
\[
\int[dk_{1}dk_{2}dk_{3}...dk_{n}...]\Psi [k_{0},k_{1},k_{2},..k_{n},..]
= \phi (k_{0})
\]
\[
\int[dk_{1}dk_{2}dk_{3}...dk_{n}...]k_{1}^{\mu}
\Psi [k_{0},k_{1},k_{2},..k_{n},..]
= A_{\mu}(k_{0})
\]
\[
\int[dk_{1}dk_{2}dk_{3}...dk_{n}...]k_{1}^{\mu}k_{1}^{\nu}
\Psi [k_{0},k_{1},k_{2},..k_{n},..]
=S^{\mu \nu}(k_{0} )
\]
\[
\int[dk_{1}dk_{2}dk_{3}...dk_{n}...]k_{2}^{\mu}
\Psi [k_{0},k_{1},k_{2},..k_{n},..]
=S^{\mu} (k_{0} )
\]
Here, $\Psi$ is the string field and $\phi$,$A_{\mu} \, ,S^{\mu \nu}, \,
S^{\mu}$ are the tachyon, the massless vector and two massive modes.
\footnote{The auxiliary fields are also included.}

If we define $\alpha _{i}$ by
\be     \label{3.12}
\alpha (t) = \al _{0} +
\frac{\al _{1}}{t}
+ \frac{\al _{2}}{t^{2}}
+...+\frac{\al _{n}}{t^{n}} +...
\ee
then
\be     \label{3.13}
Y \equiv X+\al _{1} \p X+ \al _{2} \pp X + \frac{\al _{3} \ppp X}{3!} +...
+ \frac{\al _{n} \p ^{n} X}{(n-1)!} +...
\ee
Furthermore
\be     \label{3.14}
Y_{i} \equiv \frac{\p Y} {\p x_{i}}
\ee
where $x_{i}$ are defined by
\be     \label{3.15}
\al (t) \equiv e^{\sum _{i}x_{i}t^{-i}}
\ee
Thus they satisfy
\be     \label{3.16}
\frac{\p \al _{n}}{\p x_{i}} = \al _{n-i}
\ee
The $x_{n}$ are an infinite number of variables that describe
reparametrizations of the boundary of the world sheet on which the
loop variables are defined.  In the Polyakov formalism these
have to be integrated over.  Thus the loop variables come with
$\int [dx_{n}]$ attached to them.  The gauge transformation
on the string field is summarized by
\be     \label{3.17}
k(t) \rightarrow k(t) \lambda (t)
\ee
This completes our review.  For full details see \cite{bslv,bslv1}.

We would like to use these variables in our generalized proper time
formalism.  We thus need to know the action of $\int dt \lambda (t)
T_{zz}(z+t)$ on the loop variable (\ref{3.10}).  In fact the answer
to this is known in closed form \cite{bsvg}. If we use the notation
\footnote{In \cite{bsvg} we used the notation $Y_{m}$ for this.
But we have already used $Y_{m}$ for $\frac{\partial Y}{\partial x_{m}}$.
If $\alpha (t) =1$ then $Y_{m} = \tilde{Y}_{m}$.}
\be     \label{3.18}
\tilde{Y} \equiv \frac{\p ^{m}X}{(m-1)!}
\ee
then
\be     \label{3.19}
\sum _{n} k_{n} Y_{n} = \sum _{m}K_{m} \tilde{Y}_{m}
\ee
where
\be     \label{3.20}
K_{m} \equiv k_{0}\al _{m} +
k_{1}\al _{m-1}+
k_{2}\al _{m-2}+...+k_{m}
\ee
Thus
\be     \label{3.21}
\gvk = e^{i\sum _{m} K_{m} \tilde{Y}_{m}}
\ee
{}From \cite{bsvg} we know the exact expression for the operator product
expansion
$\eln e^{iK_{m}\tilde{Y}_{m}}$.  Here we need it only to linear order,
and only for $n\geq 0$.  The result is:
\be     \label{3.22}
\eln :e^{iK_{m}\tilde{Y}_{m}}: =
e^{-pq\lambda _{-p-q}K_{p}.K_{q}}:e^{imK_{m}\lambda _{n-m}\tilde{Y}_{n}}
e^{iK_{m}\tilde{Y}_{m}+O(\lambda ^{2})}:
\ee
The first factor is the ``quantum'' or ``anomalous'' piece, whereas
the second one is the ``classical'' piece.  The latter can be ignored
for our purposes since it is like a field redefinition \cite{bslv}.
Applying (\ref{3.22}) to (\ref{3.10}) we get (using (\ref{3.20}))
\be     \label{3.23}
\eln :\gvk : = e^{-\sum _{n,m,p}k_{n}.k_{m}
[\sum _{q}q(p-q)\alpha _{q-n} \alpha _{p-m-q}]\lambda _{-p}}: \gvk :
\ee
Let us define a field $\sigma$, a linear function of all the
$\lambda _{-p}$ and $x_{n}$ by:
\be     \label{3.24}
\sigma \equiv
\sum _{q}\sum _{p}q(p-q)\alpha _{q} \alpha _{p-q}\lambda _{-p}
\ee
Then one can show that the RHS (\ref{3.23}) can be rewritten
as
\be     \label{3.25}
 e^{-\sum _{n,m}\frac{k_{n}.k_{m}}{2}
[\dsnm -\dsq ]}: \gvk :
\ee
In this form the result is exactly that of \cite{bslv}.  $\sigma$ can
be identified with the ``new'' Liouville mode introduced there.
If $\alpha (t)=1$ the ``new'' mode is just the usual Liouville mode and
in fact (\ref{3.25}) gives the Liouville mode dependence of a generalized
vertex operator due to the Weyl anomaly.

We can now use all this in the proper time equation
\be     \label{3.26}
\dds \int [dx_{n}]
\left. <\oint dt \lambda (t) T_{zz} (z+t) V_{I}(z) V_{J}(0)> \right |
_{\sigma =0} \Phi ^{I} (0) = \, 0
\ee
In evaluating (\ref{3.26}) we will use a two point function
\be     \label{3.27}
-\Sigma (z) \equiv <Y(z)Y(0)> \, = \, \alpha '
(\ln z + O(\alpha _{1} z^{-1}.. .)
\ee
Here we have introduced the string tension $\alpha '$ to emphasize
that higher orders in $\Sigma$ are also higher orders in $\alpha '$.
When (\ref{3.26}) is evaluated we will keep only terms linear in
$\Sigma$.  Furthermore we have loop variables both at $z$ and at $0$, and
each has a set of variables that describe reparametrizations.
We denote the ones at $z$ by $x_{1}, x_{2}...$ and the ones at $0$
by $y_{1},y_{2},..$.  Thus $\Sigma $ is a function of both $x_{n}$
and $y_{n}$.

Now, if we consider the following vertex operator and its associated
$\sigma $ dependence,
\be     \label{3.28}
e^{i(k_{0} Y +k_{1}Y_{1} + k_{2}Y_{2}) - k_{0} ^{2} \sigma -
\frac{k_{1}.k_{1}}{2}
(\dsii - \dst ) - k_{2}.k_{0} \dst -k_{1}k_{0} \dsi }
\ee
we have all we need to go to the second mass level.

For the photon we get:
\be     \label{3.29}
\left.\int dx_{1} \dds <(-k_{1}. k_{0} \dsi +
ik_{1}\frac{\p Y}{\p x_{1}})
e^{ik_{0} Y(z) - k_{0} ^{2} \sigma }iq_{1}
\frac{\p Y}{\p y_{1}}e^{iq_{0}Y(0)}>\right | _{\sigma =0}
=0
\ee
\be     \label{3.30}
\Rightarrow  \left.
\int dx_{1} \dds (-k_{1}.k_{0} \dsi k_{0} . q_{1}\frac{\p \Sigma }
{\p y_{1}}
+k_{1}. q_{1}
\frac{\pp \Sigma }{\p x_{1} \p y_{1}}
)e^{k_{0} .q_{0}\Sigma
- k_{0} ^{2} \sigma} \right | _{\sigma =0}=\, 0
\ee
\be     \label{3.31}
\Rightarrow
(k_{1}. k_{0} k_{0} . q_{1}- k_{0} ^{2}k_{1}. q_{1})
\frac{\pp \Sigma }{\p x_{1} \p y_{1}} =0
\ee
The coefficient of $q_{1}^{\mu}$ is the equation of motion for the photon.

We can also apply this very easily to the next mass level.  Since the
calculation is very similar to that of \cite{bslv,bslv1} we will
merely write down the proper time equation and give a few results:
The equation is
\[
\dds \int dx_{1}dx_{2}< (\frac{1}{2}k_{1}.k_{1}(\dsii - \dst )
-k_{2}. k_{0} \dst -ik_{1}. k_{0} \dsi \kim
\frac{\p Y^{\mu}}{\p x_{1}}
\]
\[
+ i \ktm
\frac{\p Y^{\mu}}{\p x_{2}}
-\frac{\kim \kin}{2}
\frac{\p Y^{\mu}}{\p x_{1}}\frac{\p Y^{\nu}}{\p x_{1}})
e^{ik_{0} Y(z) - k_{0} ^{2}
\sigma }
\]
\be     \label{3.32}
\left.
e^{il_{0}Y(0)} (il_{2}^{\rho}
\frac{\p Y^{\rho}}{\p y_{2}} -
\frac{l_{1}^{\rho}l_{1}^{\sigma}}{2}
\frac{\p Y^{\rho}}{\p y_{1}}\frac{\p Y^{\sigma}}{\p y_{1}})>
\right | _{\sigma =0} = \, 0
\ee
One can evaluate each contraction and keep only those linear in $\Sigma$.
The coefficient of $l_{2}^{\mu}$ gives the equation for the massive
spin 1 (auxiliary) field $S^{\mu}$ and that of $l_{1}^{\mu}l_{1}^{\nu}$
gives the equation for the massive spin 2 field $S^{\mu \nu}$.
We will only give the equation for $S^{\mu}$:
\be     \label{3.33}
[-k_{1}.k_{1}\kom + k_{2}. k_{0} \kom +k_{1}. k_{0} \kim - k_{0} ^{2} \ktm ]
\frac{\pp \Sigma }{\p x_{2} \p y_{2}}=0
\ee
One can check that it is invariant under
\be     \label{3.34}
k_{2}\rightarrow k_{2}+ \la _{1}k_{1}+ \la _{2} k_{0} \, \, \, \,
k_{1}\rightarrow k_{1}+ \la _{1} k_{0}
\ee
This is all exactly as in \cite{bslv1}.  As shown there,
the massive equations
can be obtained by dimensional reduction and some identifications
of the $k_{i}$ with each other to reduce the number of degrees of
freedom to that of conventional string theory.  We will not repeat the
details here.

What we have done so far is to generalize the proper time equation
for higher modes of the string and we have done this by
incorporating the loop variable approach of \cite{bslv,bslv1}.
We can now combine this with the results of the previous section
and write down the minimal coupling of electromagnetism to
massive modes.  However this violates the gauge invariances
(\ref{3.34}) and is not consistent.  To restore consistency
one has to include the interaction of the $F_{\mu \nu}$
part in (\ref{id}).  \footnote{The full coupling has been worked out
in the zero momentum limit using string field theory in \cite{an}.}
In the general momentum
dependent case, this can be done by a perturbative evaluation of the proper
time equation.  In order to do this one has to integrate over
Koba-Nielsen parameters.  But the presence of $x_{n}$ complicates
matters.  In the BRST formalism, instead
of $x_{n}$, there are ghost fields - but these are amenable to standard
conformal field theory techniques.  Whether something analogous
is possible here is an open question that requires further study.

\newpage
\section{Conclusions}

In this paper two things have been done. First, we have introduced
the equivalent of a covariant derivative and explained how minimal
coupling of gauge fields can be implemented in a natural way.  This
captures the geometric idea of the gauge field being a connection.
This kind of facility is perhaps one of the main attractions
of the $\sigma $-model approach.

Second,
the proper time equation has been generalized, by incorporating
loop variables, to deal with higher spin modes.  The formalism looks very
similar to the background independent formalism of Witten \cite{w,wl,s}.
For massive modes we have done some explicit calculations at the free level.

The issue of interactions, in the case of massive modes, is not
fully resolved.
\footnote{The massless case has been worked out in \cite{bsz}.}
It is possible to introduce a minimal interaction with gauge fields,
but by itself this is not consistent since it is not gauge invariant.
As explained at the end of the last
section, the crux of the problem is to understand how to do the
Koba-Nielsen integrals in the presence of the $x_{n}$'s.  We hope
to return to this problem soon.


\newpage
\begin{thebibliography}{99}
\bibitem{l}C. Lovelace, Phys. Lett.135B(1984) 75.
\bibitem{cc}C. G. Callan, D. Friedan, E. Martinec and M. Perry, Nucl.
Phys. B262(1985)593.
\bibitem{as}A. Sen, Phys. Rev. D32 (1985) 2102.
\bibitem{ft}E. Fradkin and A. A. Tseytlin, Phys. Lett.15B(1985)316.
\bibitem{cz}C. G. Callan and Z. Gan, Nucl. Phys. B272(1987)647.
\bibitem{ds}S. Das and B. Sathiapalan, Phys. Rev. Lett. 56(1986)2664.
\bibitem{ao} R. Akhoury and Y. Okada, Phys. Lett. 183B (1985)65.
\bibitem{p}J. Hughes, J. Liu, and J. Polchinski, Nucl. Phys. B316(1989)
\bibitem{bm}T. Banks and E. Martinec, Nucl. Phys. B294 (1987)733.
\bibitem{bspt}B. Sathiapalan, Nucl. Phys. B294(1987)747.
\bibitem{ks}I. Klebanov and L. Susskind, Phys. Lett. B200 (1988) 446.
\bibitem{bny}R. Brustein, D. Nemeschansky and S. Yankielowicz,
Nucl. Phys. B301 (1988) 224.
\bibitem{ef}U. Ellwanger and J. Fuchs,Nucl. Phys. B312 (1989) 95.
\bibitem{bslv}B. Sathiapalan, Nucl. Phys. B326(1989)376.
\bibitem{bslv1}B. Sathiapalan,Nucl. Phys. B415(1994)332.
\bibitem{ac}A. Abouelsaood, C. G. Callan, C. R. Nappi and S. A. Yost,
Nucl. Phys. B280 (1989) 599.
\bibitem{ft1} E.S. Fradkin and A. A. Tseytlin , Phys. Lett. B163(1985)
123.
\bibitem{t}  A. A. Tseytlin , Nucl. Phys. B276 (1986)391.
\bibitem{do} H. Dorn and H. J. Otto, Zeit. f. Phys. C32(1986)599.
\bibitem{bsos}B. Sathiapalan, Phys. Lett. B201(1988)454.
\bibitem{ft2}E. S. Fradkin and A. A. Tseytlin, Phys Lett. 158B(1985)316.
\bibitem{t2}A. A. Tseytlin, Phys. Lett. 185B, (1987) 59; 168B(1983)63.
\bibitem{bil}A. Bilal, hepth 9508062, LPTENS-95/32.
\bibitem{bsrg}B. Sathiapalan, hepth 9505134,Mod. Phys. Lett A10 (1995) 1565.
\bibitem{bsz}B. Sathiapalan, hepth 9509097.
\bibitem{bsg}B. Sathiapalan, Mod. Phys. Lett. A9(1994) 1681.
\bibitem{t3}A. A. Tseytlin, hepth 9505052 and references therein.
\bibitem{ds2}S. R. Das and B. Sathiapalan, Phys. Rev. Lett. 57(1986) 1511.
\bibitem{amit}D. J. Amit, Y. G. Goldschmidt, and G. Grinstein, J. Phys.A13,
(1980) 585.
\bibitem{bsfc} B. Sathiapalan, hepth9409023, to appear in Int. J. of
Mod. Phys.
\bibitem{sz}W. Siegel and B. Zwiebach, Nucl. Phys.B263 (1986)105.
\bibitem{bp}T. Banks and M. Peskin, Nucl. Phys.B264(1986)513.
\bibitem{w2}E. Witten, Nucl. Phys. B268 (1986) 513.
\bibitem{lpp}A. LeClair, M. Peskin and C. Preitschopf, Nucl. Phys. B317
(1989)411.
\bibitem{gj}D. Gross and A. Jevicki, Nucl. Phys. B283 (1987) 1;
Nucl. Phys. B287 (1987) 225.
\bibitem{z}B. Zwiebach, Lectures at Les Houches Summer School, hepth 9305026.
\bibitem{an}P. C. Argyres and C. Nappi, Proceedings of 1989 ICTP
Summer School.
\bibitem{bsvg}B. Sathiapalan, Nucl. Phys. B405(1993)367.
\bibitem{w}E. Witten, Phys. Rev.D46(1992)5467; Phys. Rev.D47(1993)3405.
\bibitem{wl}E. Witten and K. Li, Phys. Rev. D48(1993)860.
\bibitem{s}S. Shatashvili, Phys. Lett. 311B(1993)83.
\bibitem{zam} A. B. Zamolodchikov, JETP Lett. 43 (1986) 730.
\end{thebibliography}
\end{document}